\documentclass[twocolumn,english,pra, aps, twocolumn, superscriptaddress]{revtex4-1}
\usepackage[T1]{fontenc}
\usepackage[latin9]{inputenc}
\usepackage{xcolor}
\usepackage{pdfcolmk}
\usepackage{amsmath}
\usepackage{amssymb}
\usepackage{graphicx}
\PassOptionsToPackage{normalem}{ulem}
\usepackage{ulem}

\makeatletter
\makeatletter
\usepackage[colorlinks,citecolor=blue,linkcolor=blue]{hyperref}
\newcommand{\lyxdot}{.}

\providecolor{lyxadded}{rgb}{0,0,1}
\providecolor{lyxdeleted}{rgb}{1,0,0}

\DeclareRobustCommand{\lyxsout}[1]{\ifx\\#1\else\sout{#1}\fi}

\makeatother

\usepackage{babel}
\begin{document}
\title{Self-consistency of optimizing finite-time Carnot engines with the
low-dissipation model}
\author{Yu-Han Ma}
\address{Graduate School of China Academy of Engineering Physics, No. 10 Xibeiwang
East Road, Haidian District, Beijing, 100193, China}
\author{C. P. Sun}
\address{Graduate School of China Academy of Engineering Physics, No. 10 Xibeiwang
East Road, Haidian District, Beijing, 100193, China}
\address{Beijing Computational Science Research Center, Beijing 100193, China}
\author{Hui Dong}
\email{hdong@gscaep.ac.cn}

\address{Graduate School of China Academy of Engineering Physics, No. 10 Xibeiwang
East Road, Haidian District, Beijing, 100193, China}
\begin{abstract}
The efficiency at the maximum power (EMP) for finite-time Carnot engines
established with the low-dissipation model, relies significantly on
the assumption of the inverse proportion scaling of the irreversible
entropy generation $\Delta S^{(\mathrm{ir})}$ on the operation time
$\tau$, i.e., $\Delta S^{(\mathrm{ir})}\propto1/\tau$. The optimal
operation time of the finite-time isothermal process for EMP has to
be within the valid regime of the inverse proportion scaling. Yet,
such consistency was not tested due to the unknown coefficient of
the $1/\tau$-scaling. In this paper, using a two-level atomic heat
engine as an illustration, we reveal that the optimization of the
finite-time Carnot engines with the low-dissipation model is self-consistent
only in the regime of $\eta_{\mathrm{C}}\ll1$, where $\eta_{\mathrm{C}}$
is the Carnot efficiency. In the large-$\eta_{\mathrm{C}}$ regime,
the operation time for EMP obtained with the low-dissipation model
is not within the valid regime of the $1/\tau$-scaling, and the exact
EMP is found to surpass the well-known bound $\eta_{+}=\eta_{\mathrm{C}}/(2-\eta_{\mathrm{C}})$.
\end{abstract}
\maketitle

\section{Introduction}

Converting heat into useful work, heat engine lies at the core of
thermodynamics, both in classical and quantum regime \citep{Carnoteff,tolman1948irreversible,kosloff2014quantum,Binder2018}.
Absorbing heat from a hot thermal bath with the temperature $T_{\mathrm{h}}$,
the engine outputs work and releases part of the heat to the cold
bath with the temperature $T_{\mathrm{c}}$. The upper limit of the
heat engine working between two heat baths is given by the Carnot
efficiency $\eta_{\mathrm{C}}=1-T_{\mathrm{c}}/T_{\mathrm{h}}$ \citep{Carnoteff}.
Due to the limitation of the quasi-static cycle with infinite-long
operation time, the heat engine with Carnot efficiency generally has
vanishing output power and in turn is of no practical use. To design
the heat engine cycles operating in finite time, several practical
heat engine models have been proposed \citep{andresen1984thermodynamics,wu1999recent,ZCTuCPB},
such as the endo-reversible model \citep{reitlinger1929utilisation,yvon1955saclay,chambadal1957recuperation,novikov1958efficiency,CA},
the linear irreversible model \citep{BroeckPRL2005,Wang2012,izumida2014work},
the stochastic model \citep{schmiedl2008efficiency,StochasticThermodynamics},
and the low-dissipation model \citep{EspositoPRL2010,2013EfficiencyBroeck,2015Efficiency,Constraintrelationyhma,2020LDmodel,2020Optimal}.
The efficiency at maximum power (EMP), is proposed as an important
parameter to evaluate the performance of these heat engines in the
finite-time cycles.

The utilization of the low-dissipation model\citep{EspositoPRL2010,2013EfficiencyBroeck,2015Efficiency,Constraintrelationyhma,yhmaoptimalcontrol}
simplifies the optimization of the finite-time Carnot-like heat engines.
As the model assumption, the heat transfer between the engine and
the bath in the finite-time quasi-isothermal process is divided into
two parts as follow

\begin{equation}
Q_{\mathrm{h,c}}(\tau_{\mathrm{h}})=T_{\mathrm{h,c}}(\Delta S_{\mathrm{h,c}}-S_{\mathrm{h,c}}^{(\mathrm{ir})}),\label{eq:Qhc}
\end{equation}
where $\Delta S_{\mathrm{h}}=-\Delta S_{\mathrm{c}}=\Delta S$ is
the reversible entropy change of the working substance and $S_{\mathrm{h,c}}^{(\mathrm{ir})}=\Sigma_{\mathrm{h,c}}/\tau_{\mathrm{h,c}}$
is the irreversible entropy generation which is inversely proportional
to the process time $\tau_{\mathrm{\alpha}}$. Optimizing the output
power $P(\tau_{\mathrm{h}},\tau_{\mathrm{c}})=[Q_{\mathrm{h}}(\tau_{\mathrm{h}})+Q_{\mathrm{c}}(\tau_{\mathrm{c}})]/(\tau_{\mathrm{h}}+\tau_{\mathrm{c}})$
with respect to the operation time $\tau_{\mathrm{h}}$ and $\tau_{\mathrm{c}}$,
one gets the optimal operation times \citep{EspositoPRL2010} as

\begin{align}
\tau_{\mathrm{h}}^{*} & =\frac{2T_{\mathrm{h}}\Sigma_{\mathrm{h}}}{(T_{\mathrm{h}}-T_{\mathrm{c}})\Delta S}\left(1+\sqrt{\frac{T_{\mathrm{c}}\Sigma_{\mathrm{c}}}{T_{\mathrm{h}}\Sigma_{\mathrm{h}}}}\right),\label{eq:th}\\
\tau_{\mathrm{c}}^{*} & =\frac{2T_{\mathrm{c}}\Sigma_{\mathrm{c}}}{(T_{\mathrm{h}}-T_{\mathrm{c}})\Delta S}\left(1+\sqrt{\frac{T_{\mathrm{h}}\Sigma_{\mathrm{h}}}{T_{\mathrm{c}}\Sigma_{\mathrm{c}}}}\right),\label{eq:tc}
\end{align}
and the efficiency at the maximum power $\eta^{*}$ bounded by the
following inequality as \citep{EspositoPRL2010,ZCTuCPB}

\begin{equation}
\eta_{\mathrm{-}}\equiv\frac{\eta_{\mathrm{C}}}{2}\leq\eta^{*}\leq\frac{\eta_{\mathrm{C}}}{2-\eta_{\mathrm{C}}}\equiv\eta_{\mathrm{+}}.\label{eq:inequality}
\end{equation}
Due to the simplicity of the model assumption and the universality
of the obtained EMP, the low-dissipation model becomes one of the
most studied finite-time heat engine models in recent years \citep{2013EfficiencyBroeck,2015Efficiency,Constraintrelationyhma,2020LDmodel,2020Optimal,2020-finite-size}.

It is currently cleared that \citep{TradeoffrelationShiraishi,CavinaPRLtradeoffrelation,Constraintrelationyhma,yhmaoptimalcontrol,2020IEGyhma}
the low-dissipation assumption is valid in the long-time regime of
$\tau_{\mathrm{h(c)}}/t_{\mathrm{r}}\gg1$, where $t_{\mathrm{r}}$
is the relaxation time for the work substance to reach its equilibrium
with the heat bath. And the dissipation coefficient $\Sigma_{\mathrm{h(c)}}$
of the $1/\tau$-scaling is determined by both the coupling strength
$\gamma_{\mathrm{h(c)}}\sim1/t_{\mathrm{r}}$ to the bath\citep{Constraintrelationyhma,2020IEGyhma}
and the control scheme \citep{yhmaoptimalcontrol,2020IEGyhma}. Such
a relation implies that the condition $\tau_{\mathrm{h(c)}}^{*}/t_{\mathrm{r}}\gg1$
is not fulfilled simply and should be justified to reveal the regime
of validity. We check the consistency of the obtained EMP with a minimal
heat engine model consisting of a single two-level system. In Sec.
\ref{sec:Self-consistency-of-low-dissipat}, we analytically obtain
the regime, where the optimal operation time to achieve EMP is consistent
with the low-dissipation assumption. And we further show the possibility
of the exact EMP of the engine to surpass the upper bound of EMP,
i.e., $\eta_{\mathrm{+}}$, obtained with the low-dissipation model
in the large-$\eta_{\mathrm{C}}$ regime in Sec. \ref{sec:Efficiency-at-maximum}.

\section{\label{sec:Self-consistency-of-low-dissipat}Self-consistency of
the low-dissipation model in deriving efficiency at maximum power}

The two-level atomic heat engine is the simplest quantum engine to
demonstrate the relevant physical mechanisms \citep{Geva1992,QTquanQH,Su2018,Constraintrelationyhma,yhmaoptimalcontrol}.
The energy spacing of the excited state $\left|e\right\rangle $ and
ground state $\left|g\right\rangle $ is tuned by an outside agent
to extract work with the Hamiltonian $H=\frac{1}{2}\hbar\omega\left(t\right)\sigma_{z},$
where $\sigma_{z}=\left|e\right\rangle \left\langle e\right|-\left|g\right\rangle \left\langle g\right|$
is the Pauli matrix in the z-direction. The Planck's constant is taken
as $\hbar=1$ in the following discussion for convenience. For the
finite-time quasi-isothermal process with the duration $\tau$ of
the two-level system, the low-dissipation assumption of the $1/\tau$
scaling is valid in the regime $\widetilde{\gamma}\tau\gg1$ \citep{Constraintrelationyhma},
where $\widetilde{\gamma}=2\gamma T/\omega_{0}$. Here $\gamma$ is
the coupling strength between the system and the bath with the temperature
$T$ and $\omega_{0}$ is the initial energy spacing of the system
during the process.

\begin{figure}
\centering{}\includegraphics[width=8.5cm]{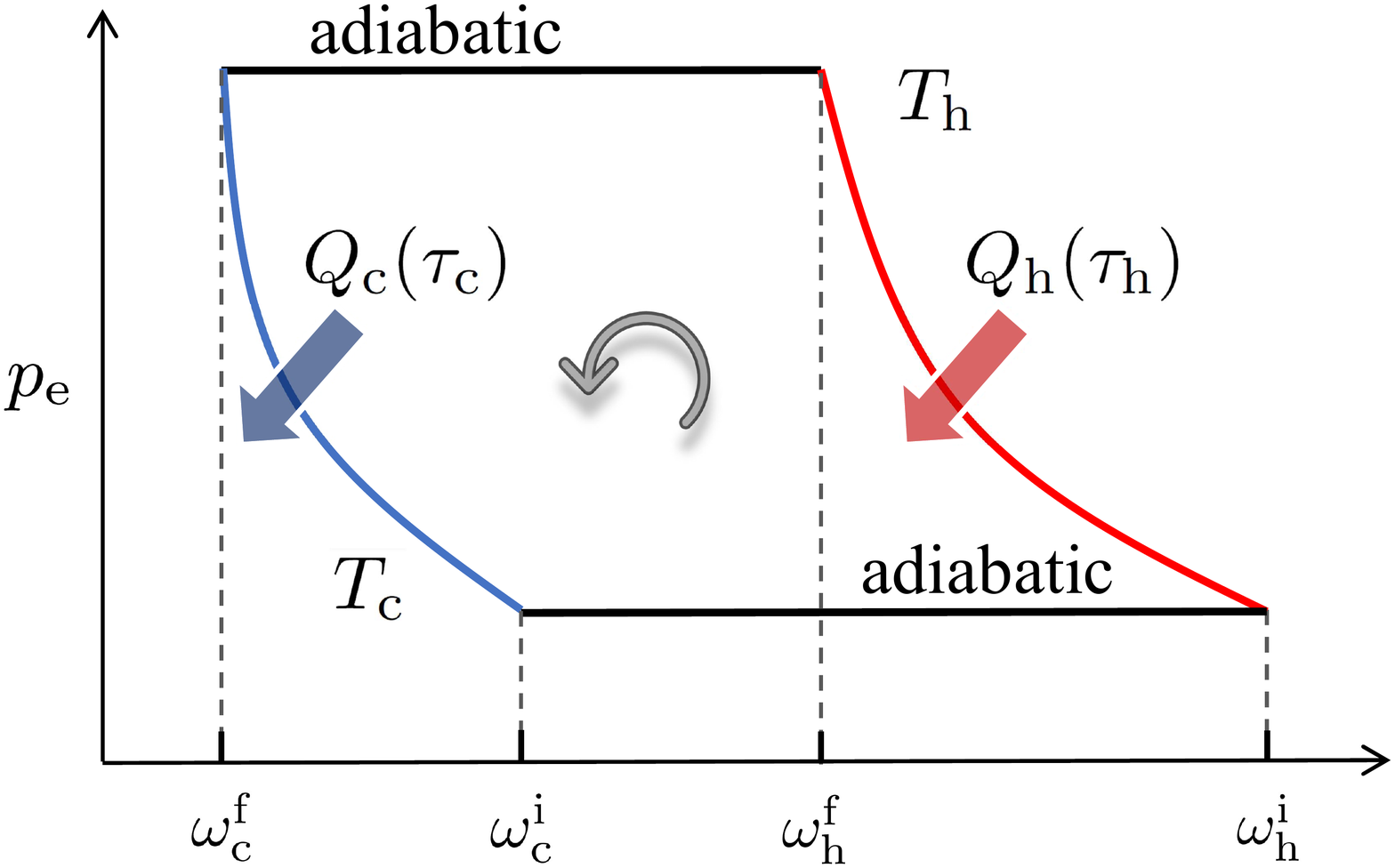}\caption{\label{fig:Schematic-diagram-of}Schematic diagram of the finite-time
Carnot-like cycle for a two-level atomic heat engine. The horizontal
axis and the vertical axis represent respectively the energy spacing
$\omega$ and excited state population $p_{\mathrm{e}}$ of the two-level
atom. The red (blue) solid curve represents the high (low) temperature
finite-time quasi-isothermal process, and the black solid lines represent
the adiabatic processes.}
\end{figure}

The finite-time Carnot-like cycle for the two-level atomic heat engine
of interest consists of four strokes, two isothermal and two adiabatic
processes. The schematic diagram of the cycle is shown in Fig. \ref{fig:Schematic-diagram-of}.
In the figure, $\omega_{\mathrm{h}}^{\mathrm{i}}$ and $\omega_{\mathrm{h}}^{\mathrm{f}}$
($\omega_{\mathrm{c}}^{\mathrm{i}}$ and $\omega_{\mathrm{c}}^{\mathrm{f}}$
) are respectively the initial and finial energy spacing of the working
substance in the high (low) temperature finite-time quasi-isothermal
process with duration $\tau_{\mathrm{h}}$ ($\tau_{\mathrm{c}}$),
which is shown with the red (blue) solid curve. The total operating
time per cycle is $t=\tau_{\mathrm{h}}+\tau_{\mathrm{c}}$. The interval
of the adiabatic processes (the black solid lines) are ignored in
comparison with $\tau_{\mathrm{h}}$ and $\tau_{\mathrm{c}}$ \citep{EspositoPRL2010,Constraintrelationyhma}.
We assume the two-level system has no energy level crossing during
the whole cycle to ensure no coherence of the system is induced by
non-adiabatic transition\citep{2012Quantum-non-adiabatic,2018Time-non-adiabatic}.
The quasi-isothermal process retains the normal isothermal process
at the quasi-static limit of $\tau_{\mathrm{h(c)}}\rightarrow\infty$.

For simplicity, we focus on the high-temperature regime, where the
reversible entropy change $\Delta S_{\alpha}$ and the irreversible
entropy generation coefficient $\Sigma_{\alpha}$ in Eq. (\ref{eq:Qhc})
are analytically written as \citep{Constraintrelationyhma}

\begin{equation}
\Delta S=\frac{\left[\left(\omega_{\mathrm{\alpha}}^{\mathrm{i}}\right)^{2}-\left(\omega_{\mathrm{\alpha}}^{\mathrm{f}}\right)^{2}\right]}{8T_{\mathrm{\alpha}}^{2}},\Sigma_{\alpha}=\frac{2\Delta S}{\widetilde{\gamma}_{\alpha}},\label{eq:deltaS-sigma}
\end{equation}
with $\alpha=\mathrm{h,c}$, and $\widetilde{\gamma}_{\alpha}=2\gamma_{\alpha}T_{\alpha}/\omega_{\mathrm{\alpha}}^{\mathrm{i}}$.
Here and after, the Boltzmann's constant $k_{\mathrm{B}}=1$ is chosen.
To obtain the above equations, the relations $\omega_{\mathrm{h}}^{\mathrm{i}}/T_{\mathrm{h}}=\omega_{\mathrm{c}}^{\mathrm{f}}/T_{\mathrm{c}}$,
$\omega_{\mathrm{h}}^{\mathrm{f}}/T_{\mathrm{h}}=\omega_{\mathrm{c}}^{\mathrm{i}}/T_{\mathrm{c}}$
have been used in the quantum adiabatic processes\citep{QTquanQH,Constraintrelationyhma}.
Substituting Eq. (\ref{eq:deltaS-sigma}) into Eqs. (\ref{eq:th})
and (\ref{eq:tc}), we obtain the corresponding optimal operation
time $\tau_{\mathrm{\alpha}}^{*}$ for achieving the maximum power
with the dimensionless time $\widetilde{\tau}_{\alpha}^{*}\equiv\tau_{\mathrm{\alpha}}^{*}\widetilde{\gamma}_{\alpha}$
as

\begin{equation}
\widetilde{\tau}_{\mathrm{h}}^{*}=\frac{2}{\eta_{\mathrm{C}}}\frac{\omega_{\mathrm{h}}^{\mathrm{i}}-\omega_{\mathrm{h}}^{\mathrm{f}}}{\omega_{\mathrm{h}}^{\mathrm{i}}+\omega_{\mathrm{h}}^{\mathrm{f}}}\left[1+\sqrt{\left(1-\eta_{\mathrm{C}}\right)\frac{\gamma_{\mathrm{h}}}{\gamma_{\mathrm{c}}}}\right],\label{eq:th-1}
\end{equation}

\begin{equation}
\widetilde{\tau}_{\mathrm{c}}^{*}=\frac{2}{\eta_{\mathrm{C}}}\frac{\omega_{\mathrm{h}}^{\mathrm{i}}-\omega_{\mathrm{h}}^{\mathrm{f}}}{\omega_{\mathrm{h}}^{\mathrm{i}}+\omega_{\mathrm{h}}^{\mathrm{f}}}\left[\sqrt{\left(1-\eta_{\mathrm{C}}\right)\frac{\gamma_{\mathrm{c}}}{\gamma_{\mathrm{h}}}}+1-\eta_{\mathrm{C}}\right].\label{eq:tc-1}
\end{equation}
The low-dissipation assumption is valid in the regime $\widetilde{\tau}_{\mathrm{h}}^{*}\gg1$
and $\widetilde{\tau}_{\mathrm{c}}^{*}\gg1$, namely,

\begin{equation}
1+\sqrt{\left(1-\eta_{\mathrm{C}}\right)\frac{\gamma_{\mathrm{h}}}{\gamma_{\mathrm{c}}}}\gg\frac{\eta_{\mathrm{C}}}{2}\frac{\omega_{\mathrm{h}}^{\mathrm{i}}+\omega_{\mathrm{h}}^{\mathrm{f}}}{\omega_{\mathrm{h}}^{\mathrm{i}}-\omega_{\mathrm{h}}^{\mathrm{f}}}\label{eq:condition1}
\end{equation}

\begin{equation}
\sqrt{\left(1-\eta_{\mathrm{C}}\right)\frac{\gamma_{\mathrm{c}}}{\gamma_{\mathrm{h}}}}+1-\eta_{\mathrm{C}}\gg\frac{\eta_{\mathrm{C}}}{2}\frac{\omega_{\mathrm{h}}^{\mathrm{i}}+\omega_{\mathrm{h}}^{\mathrm{f}}}{\omega_{\mathrm{h}}^{\mathrm{i}}-\omega_{\mathrm{h}}^{\mathrm{f}}}\label{eq:condition2}
\end{equation}
The above two inequalities are fulfilled when

\begin{equation}
\eta_{\mathrm{C}}\ll2\frac{1-\delta}{1+\delta},\label{eq:regime-1}
\end{equation}
where $\delta\equiv\omega_{\mathrm{h}}^{\mathrm{f}}/\omega_{\mathrm{h}}^{\mathrm{i}}$
is the compression ratio of the heat engine cycle in the quasi-isothermal
process. The above relation is one of the main results of the current
work and reveals the range of $\eta_{\mathrm{C}}$ in which the low-dissipation
model is applicable for finding EMP. The bound for EMP obtained in
the low-dissipation regime, as given by Eq. (\ref{eq:inequality}),
thus may be not unconditionally applicable to such two-level atomic
engine. Indeed, we will show the EMP out of the regime is larger than
the upper bound $\eta_{+}$ predicted by the low-dissipation model
in the next section.

\section{\label{sec:Efficiency-at-maximum}Efficiency at maximum power: beyond
the low dissipation model}

With the analytical discussion above, we find the EMP obtain with
the low-dissipation model is only consistent with the assumption of
the low-dissipation model in the low-$\eta_{\mathrm{C}}$ regime for
the two-level system. The question is whether the bound provided by
the low-dissipation model, i.e. $\eta_{+}$, is still the upper bound
for the achievable efficiency of the system out of the low-$\eta_{\mathrm{C}}$
regime. Unfortunately, the answer is no. In this section, we will
focus the efficiency at the maximum power in the regime with large
$\eta_{\mathrm{C}}$.

By numerically simulating the dynamics of the two-level system engine
with different cycle time, we obtain the exact power and efficiency
to find the EMP. The results in the large-$\eta_{\mathrm{C}}$ regime
show that: (i) the optimal operation time corresponding to the maximum
power of the heat engine does not meet the low-dissipation assumption;
(ii) the EMP surpass the upper bound obtained with the low-dissipation
model, namely, $\eta_{\mathrm{MP}}>\eta_{+}$.

The dynamics of the two-level atom in the finite-time quasi-isothermal
process is given by the master equation as follow \citep{Constraintrelationyhma}

\begin{equation}
\frac{dp_{\mathrm{e}}(t)}{dt}=-\kappa(t)p_{\mathrm{e}}(t)+C(t),\label{eq:master equation-1}
\end{equation}
where $p_{\mathrm{e}}(t)$ is the excited state population and $C(t)=\gamma n[\omega(t)]$.
$\kappa(t)=\gamma\left(2n[\omega(t)]+1\right)$ is the effective dissipation
rate with the mean occupation number $n[\omega(t)]=1/\left(\exp[\beta\omega(t)]-1\right)$
for the bath mode $\omega(t)$. The dissipation rate $\gamma$ equals
to $\gamma_{\mathrm{h}}$ ($\gamma_{\mathrm{c}}$) in the high (low)
temperature quasi-isothermal process with the inverse temperature
$\beta=1/(k_{\mathrm{B}}T_{\mathrm{h}})$ ($\beta=1/(k_{\mathrm{B}}T_{\mathrm{c}})$).
The energy spacing of the two-level atom is tuned linearly as $\omega(t)=\omega_{\mathrm{h}}^{\mathrm{i}}+(\omega_{\mathrm{h}}^{\mathrm{f}}-\omega_{\mathrm{h}}^{\mathrm{i}})t/\tau_{\mathrm{h}},\:t\in[0,\tau_{\mathrm{h}}]$
in the high-temperature finite-time quasi-isothermal process and as
$\omega(t)=\omega_{\mathrm{c}}^{\mathrm{i}}+(\omega_{\mathrm{c}}^{\mathrm{f}}-\omega_{\mathrm{c}}^{\mathrm{i}})t/\tau_{\mathrm{c}},\:t\in[\tau_{\mathrm{c}},\tau_{\mathrm{c}}+\tau_{\mathrm{h}}]$
in the low-temperature finite-time quasi-isothermal process. The population
of the two-level system keeps unchanged during the adiabatic processes
whose operation time is ignored in comparison with $\tau_{\mathrm{h}}$
and $\tau_{\mathrm{c}}$.

In the following simulation, we set $\gamma_{\mathrm{h}}=1$ and focus
on the regime of $\gamma_{\mathrm{c}}/\gamma_{\mathrm{h}}\rightarrow\infty$,
i.e., $\Sigma_{\mathrm{c}}/\Sigma_{\mathrm{h}}\rightarrow0$, where
the upper bound $\eta_{+}=\eta_{\mathrm{C}}/(2-\eta_{\mathrm{C}})$
of EMP of the engine is achieved according to the prediction with
the low-dissipation model\citep{EspositoPRL2010}. In this regime,
the low-temperature quasi-isothermal process approaches the isothermal
process fastly enough that the operation time $\tau_{\mathrm{c}}$
is further ignored for the optimization of the cycle's output power.
The optimization is simplified as a single parameter optimization
problem: find the maximum value $P_{\mathrm{max}}$ of the cycle's
output power with respect to $\tau_{\mathrm{h}}$, and obtain the
EMP of the engine, $\eta_{\mathrm{MP}}\equiv\eta(P=P_{\mathrm{max}})$.

\begin{figure}
\begin{centering}
\includegraphics[width=8.5cm]{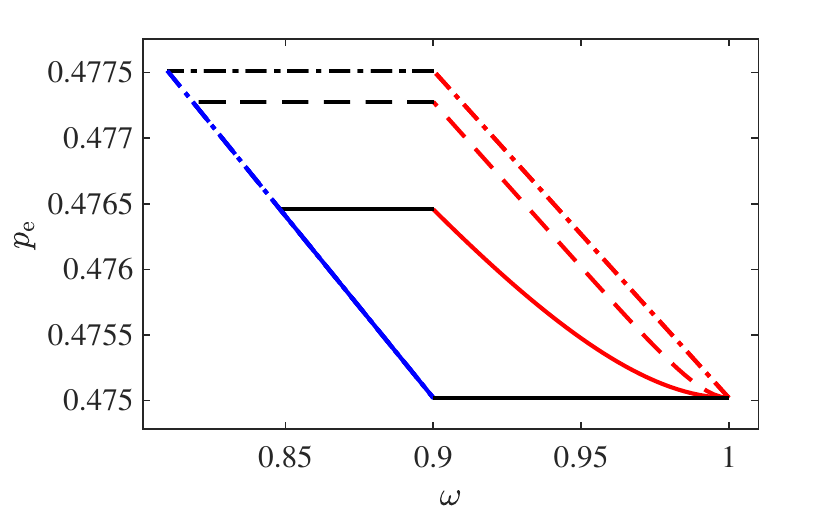}
\par\end{centering}
\centering{}\caption{\label{fig:The-finite-time-Carnot-like}The finite-time Carnot-like
cycles of for two-level atomic heat engine with different operation
time $\tau_{\mathrm{h}}$. The red curves represent the high-temperature
finite-time quasi-isothermal process with the duration $\tau_{\mathrm{h}}$,
while the blue curves represent the low-temperature isothermal process
. The adiabatic processes are plotted with the black lines. The outermost
dash-dotted curves relate to the quasi-static cycle with $\tau_{\mathrm{h}}=200t_{\mathrm{r}}$,
while the middle dashed cycle and inner solid cycle are obtained with
$\tau_{\mathrm{h}}=10t_{\mathrm{r}}$ and $\tau_{\mathrm{h}}=2t_{\mathrm{r}}$,
respectively. In this example, we choose $\omega_{\mathrm{h}}^{\mathrm{i}}=1$,
$\omega_{\mathrm{h}}^{\mathrm{f}}=0.9$, $\gamma_{\mathrm{h}}=1$,
$T_{\mathrm{h}}=10$, and $T_{\mathrm{c}}=9$. $t_{\mathrm{r}}=\omega_{\mathrm{h}}^{\mathrm{i}}/\left(2\gamma_{\mathrm{h}}T_{\mathrm{h}}\right)=0.05$
is the relaxation time related to the high-temperature finite-time
quasi-isothermal process.}
\end{figure}

The cycles with different $\tau_{\mathrm{h}}$ are illustrated in
Fig. \ref{fig:The-finite-time-Carnot-like}, where $\omega_{\mathrm{h}}^{\mathrm{i}}=1$
and $\omega_{\mathrm{h}}^{\mathrm{f}}=0.9$ are fixed. The temperatures
for the hot and cold bath are chosen as $T_{\mathrm{h}}=10$ and $T_{\mathrm{c}}=9$
as an example. The relaxation time is $t_{\mathrm{r}}=\omega_{\mathrm{h}}^{\mathrm{i}}/\left(2\gamma_{\mathrm{h}}T_{\mathrm{h}}\right)=0.05$.
The quasi-static cycles with $\tau_{\mathrm{h}}=200t_{\mathrm{r}}$,
$10t_{\mathrm{r}}$ and $2t_{\mathrm{r}}$ are represented by the
dash-dotted line, dashed line, and solid line, respectively. The figure
shows that the output work represented by the cycle area decreases
with $\tau_{\mathrm{h}}$.

\begin{figure}
\centering{}\includegraphics[width=8.5cm]{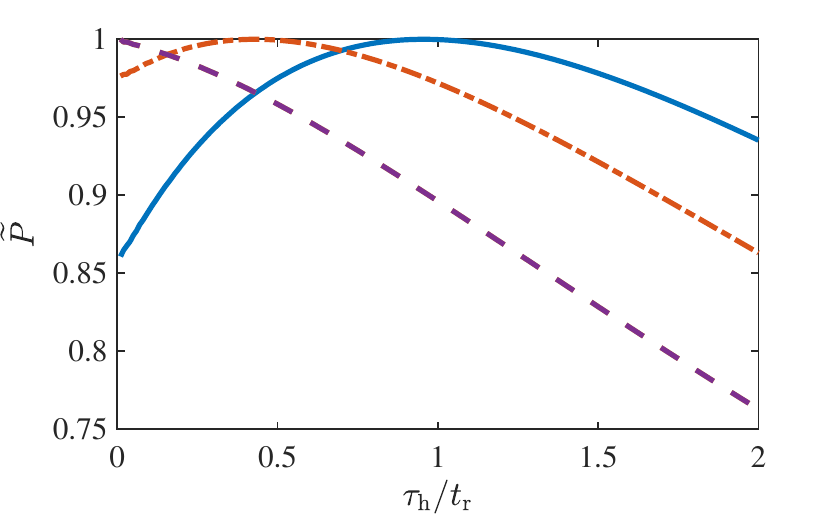}\caption{\label{fig:The-normalized-power}The normalized power of the engine
$\widetilde{P}=P/P_{\mathrm{max}}$ as the function of $\tau_{\mathrm{h}}/t_{\mathrm{r}}$.
The blue solid line, the orange dash-dotted line, and the purple dashed
line are respectively obtained with $\eta_{\mathrm{C}}=0.1$, $\eta_{\mathrm{C}}=0.12$,
and $\eta_{\mathrm{C}}=0.15$. In this example, we choose $\omega_{\mathrm{h}}^{i}=1$,
$\omega_{\mathrm{h}}^{\mathrm{f}}=0.9$, $\gamma_{\mathrm{h}}=1$,
and $T_{\mathrm{h}}=10$ with changing $T_{\mathrm{c}}=9$, $8.8$
and $8.5$. The relaxation time is $t_{\mathrm{r}}=\omega_{\mathrm{h}}^{\mathrm{i}}/\left(2\gamma_{\mathrm{h}}T_{\mathrm{h}}\right)=0.05$.}
\end{figure}

In Fig. \ref{fig:The-normalized-power}, we show the normalized power
of the engine $\widetilde{P}\equiv P/P_{\mathrm{max}}$ as the function
of $\tau_{\mathrm{h}}/t_{\mathrm{r}}$ with $\eta_{\mathrm{C}}=0.1$
(blue solid line), $\eta_{\mathrm{C}}=0.12$ (orange dash-dotted line),
and $\eta_{\mathrm{C}}=0.15$ (purple dashed line). In the simulation,
the parameters are set as $\omega_{\mathrm{h}}^{i}=1$, $\omega_{\mathrm{h}}^{\mathrm{f}}=0.9$,
and $T_{\mathrm{h}}=10$ with changing $T_{\mathrm{c}}=9$, $8.8$
and $8.5$. The relaxation time is $t_{\mathrm{r}}=0.05$. The maximum
output power $P_{\mathrm{max}}$ is obtained numerically for different
$\eta_{\mathrm{C}}$. It is observed from the figure that the dependence
of $\widetilde{P}$ on operation time $\tau_{\mathrm{h}}$ changes
with $\eta_{\mathrm{C}}$. In the figure, the optimal $\tau_{\mathrm{h}}^{*}$
decreases with $\eta_{\mathrm{C}}$ and is away from the low-dissipation
regime of $\tau_{\mathrm{h}}/t_{\mathrm{r}}\gg1$, illustrated with
the orange dash-dotted line ($\eta_{\mathrm{C}}=0.12$, $\tau_{\mathrm{h}}^{*}/t_{\mathrm{r}}\approx0.5$)
and the blue solid line ($\eta_{\mathrm{C}}=0.1$, $\tau_{\mathrm{h}}^{*}/t_{\mathrm{r}}\approx1$).
As shown clearly by the purple dashed line with $\eta_{\mathrm{C}}=0.15$,
the maximum power $\widetilde{P}=1$ is achieved in the short-time
regime of $\tau_{\mathrm{h}}/t_{\mathrm{r}}\ll1$, where the $1/\tau$-scaling
of irreversible entropy generation is invalid \citep{Constraintrelationyhma,2020IEGyhma}.

\begin{figure}
\begin{raggedright}
(a)
\par\end{raggedright}
\begin{centering}
\includegraphics[width=8.5cm]{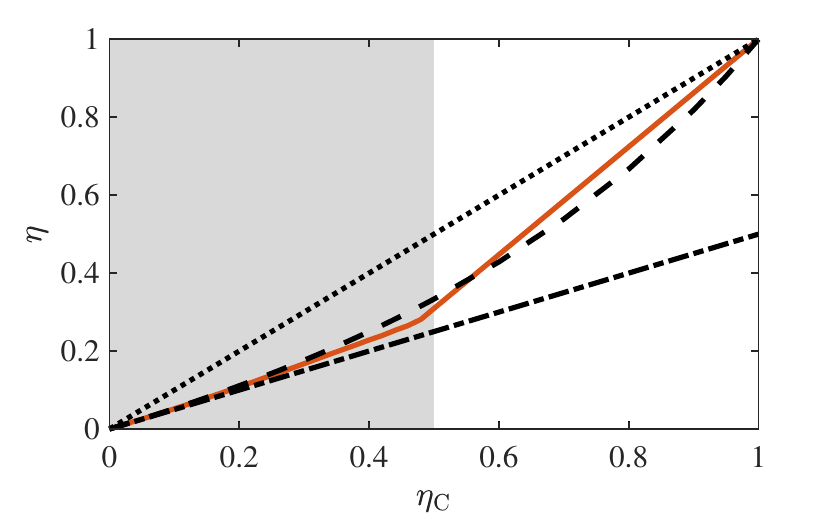}
\par\end{centering}
\begin{raggedright}
(b)
\par\end{raggedright}
\begin{centering}
\includegraphics[width=8.5cm]{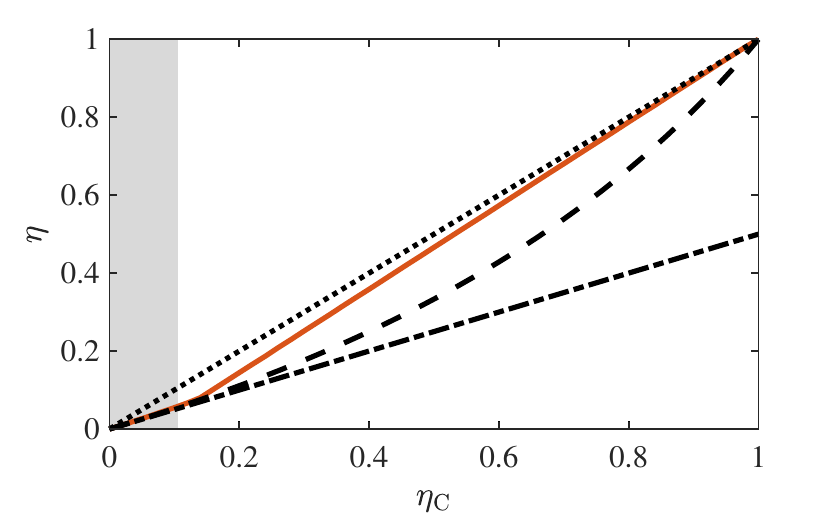}
\par\end{centering}
\begin{raggedright}
(c)
\par\end{raggedright}
\begin{centering}
\includegraphics[width=8.5cm]{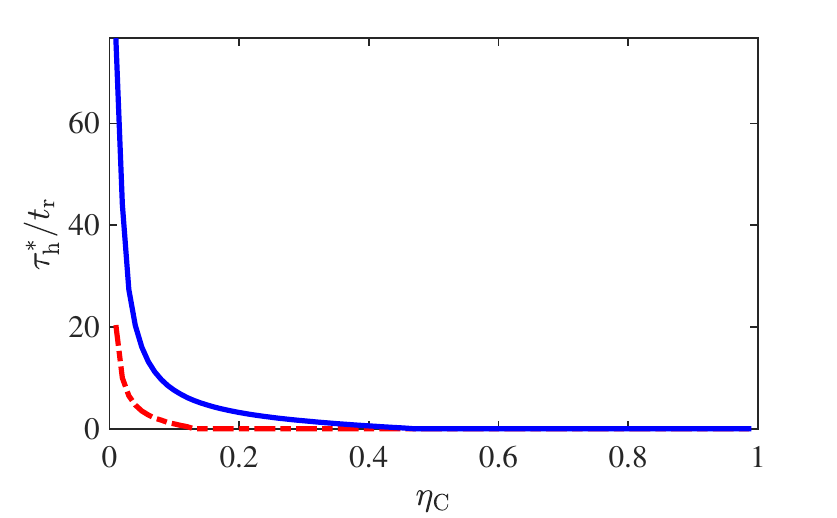}
\par\end{centering}
\raggedright{}\caption{\label{fig:Efficiency-at-maximum}Efficiency at the maximum power
$\eta_{\mathrm{MP}}$ (orange solid line) of the heat engine as the
function of the Carnot efficiency $\eta_{\mathrm{C}}$ for different
final energy spacing of the two level system(a) $\omega_{\mathrm{h}}^{\mathrm{f}}=0.6$
and (b) $\omega_{\mathrm{h}}^{\mathrm{f}}=0.9$. The black dashed
line (black dash-dotted line) represents the upper bound $\eta_{\mathrm{+}}$
(lower bound $\eta_{\mathrm{-}}$) of EMP obtained with the low-dissipation
model {[}Eq. (\ref{eq:inequality}){]}, and the Carnot efficiency
$\eta_{\mathrm{C}}$ is plotted with the black dotted line. The gray
area represents the low-dissipation regime predicted by Eq. (\ref{eq:regime-1}).
(c) Optimal operation time $\tau_{\mathrm{h}}^{*}$ at the maximum
power as the function of $\eta_{\mathrm{C}}$. The blue solid curve
is obtained with $\omega_{\mathrm{h}}^{\mathrm{f}}=0.6$ ($\delta=0.6$)
while the red dash-dotted curve is obtained with $\omega_{\mathrm{h}}^{\mathrm{f}}=0.9$
($\delta=0.9$). The other parameters in this figure are chosen as
$\omega_{\mathrm{h}}^{\mathrm{i}}=1$, $\gamma_{\mathrm{h}}=1$, and
$T_{\mathrm{h}}=10$. The relaxation time is $t_{\mathrm{r}}=\omega_{\mathrm{h}}^{\mathrm{i}}/\left(2\gamma_{\mathrm{h}}T_{\mathrm{h}}\right)=0.05$.}
\end{figure}

We show the obtained efficiency $\eta_{\mathrm{MP}}$ at the maximum
power of the engine as the function of $\eta_{\mathrm{C}}$ in Fig.
\ref{fig:Efficiency-at-maximum}(a) and (b), and plot the corresponding
optimal operation time $\tau_{\mathrm{h}}^{*}$ in Fig. \ref{fig:Efficiency-at-maximum}(c).
We choose the final energy spacing of the two level system as $\omega_{\mathrm{h}}^{\mathrm{f}}=0.6$
and $\omega_{\mathrm{h}}^{\mathrm{f}}=0.9$ respectively for (a) and
(b), and other parameters are set as $\omega_{\mathrm{h}}^{\mathrm{i}}=1$,
$\gamma_{\mathrm{h}}=1$, $T_{\mathrm{h}}=10$. As shown in Fig. \ref{fig:Efficiency-at-maximum}(a)
and (b), the EMP of the engine $\eta_{\mathrm{MP}}$(orange solid
line) in the large-$\eta_{\mathrm{C}}$ regime surpasses the upper
bound of EMP, $\eta_{+}=\eta_{\mathrm{C}}/(2-\eta_{\mathrm{C}})$
(black dashed line) obtained with the low-dissipation model. The lower
bound of EMP, $\eta_{-}=\eta_{\mathrm{C}}/2$, obtained with the low-dissipation
model is plotted with the black dash-dotted line. The gray area represents
the consistent regime as demonstrated by Eq. (\ref{eq:regime-1}).The
figure shows that $\eta_{\mathrm{MP}}$ is bounded by $\eta_{+}$
and $\eta_{-}$ of Eq. (\ref{eq:inequality}) in the gray area with
relatively small $\eta_{\mathrm{C}}$. Additionally, by comparing
(b) and (a) of Fig. \ref{fig:Efficiency-at-maximum}, with the larger
the compression rate $\delta=\omega_{\mathrm{h}}^{\mathrm{f}}/\omega_{\mathrm{h}}^{\mathrm{i}}$
($\delta=0.9$ for (a) and $\delta=0.6$ for (b)), we illustrate the
narrower the range of $\eta_{\mathrm{C}}$ in which $\eta_{\mathrm{MP}}$
is bounded by $\eta_{+}$. With the increasing of the compression
ratio $\delta$, the valid regime of optimization of the engine with
the low-dissipation model becomes smaller. And it is consistent with
the theoretical analysis of Eq. (\ref{eq:regime-1}).

In Fig. \ref{fig:Efficiency-at-maximum}(c), the optimal operation
time $\tau_{\mathrm{h}}^{*}$ at the maximum power (blue solid curve
for $\omega_{\mathrm{h}}^{\mathrm{f}}=0.6$ and red dash-dotted curve
for $\omega_{\mathrm{h}}^{\mathrm{f}}=0.9$) decreases monotonically
with increasing $\eta_{\mathrm{C}}$. The operation time at maximum
power $\tau_{\mathrm{h}}^{*}$ of the engine is not in the low-dissipation
regime of $\tau_{\mathrm{h}}/t_{\mathrm{r}}\gg1$ for the relatively
large $\eta_{\mathrm{C}}$. This explains why $\eta_{\mathrm{MP}}$
is no longer satisfies the bound provided by the low-dissipation model
in large-$\eta_{\mathrm{C}}$ regime, and verifies our analytical
analysis in Sec. \ref{sec:Self-consistency-of-low-dissipat}. In addition,
one can find in Fig. \ref{fig:Efficiency-at-maximum}(c) that the
red dash-dotted curve is lower than the blue solid curve. This leads
to a narrower parameter range of $\eta_{\mathrm{C}}$, in which the
optimal operation time $\tau_{\mathrm{h}}^{*}$ satisfies the low-dissipation
assumption, for the heat engine with $\omega_{\mathrm{h}}^{\mathrm{f}}=0.9$
than that with $\omega_{\mathrm{h}}^{\mathrm{f}}=0.6$. Therefore,
the phenomenon that the gray area in Fig. \ref{fig:Efficiency-at-maximum}(a)
is wider than that in Fig. \ref{fig:Efficiency-at-maximum}(b) is
explained from the perspective of the operation time.

\section{Conclusions and discussion}

In summary, we checked whether the optimal operation time for achieving
the maximum power is consistent with the requirement of the low-dissipation
model for the finite-time Carnot-like heat engines in this paper.
The low-dissipation model, widely used in the finite-time thermodynamics
to study EMP, relies on the assumption that the irreversible entropy
generation in the finite-time quasi-isothermal process of duration
$\tau$ follows the $1/\tau$ scaling in the long-time regime. The
operation time for the maximum power obtained from the model should
fulfill the requirement of the low-dissipation model assumption. Due
to the unknown coefficient of the $1/\tau$ scaling, the consistency
of the model in optimizing finite-time Carnot engines had not been
tested before.

In this paper, we proved that the optimal operation time for a two-level
finite-time Carnot engine achieving EMP satisfy the low-dissipation
assumption only in the low Carnot efficiency regime of $\eta_{\mathrm{C}}\ll1$.
This observation motivated us to check the EMP in the regime with
large $\eta_{\mathrm{C}}$. We calculated the EMP of the two-level
atomic heat engine in the full parameter space of $\eta_{\mathrm{C}}$.
It is found that, in the large-$\eta_{\mathrm{C}}$ regime, the true
EMP of the heat engine can surpass the upper bound for EMP, i.e.,
$\eta_{+}=\eta_{\mathrm{C}}/(2-\eta_{\mathrm{C}})$ obtained with
the low-dissipation model.

Our study on EMP in the large-$\eta_{\mathrm{C}}$ regime shall provide
a new insight for designing heat engines with better performance working
between two heat baths with large temperature difference. In addition
to affecting the EMP of the heat engine, the short-time effects caused
by fast driving may also influence the trade-off between power and
efficiency \citep{2015Efficiency,TradeoffrelationShiraishi,CavinaPRLtradeoffrelation,Constraintrelationyhma},
which needs further exploration. The predictions of this paper can
be tested on some experimental platforms \citep{BrownianHENatPhys2015,Rossnagel2016,shortcutHEexperimentSciAdv2018,2019Thermodynamic-shortcut-isothermal,2020IEGyhma,atomic-Endoreversible2020}
in the short-time regime.
\begin{acknowledgments}
This work is supported by the National Natural Science Foundation
of China (NSFC) (Grants No. 12088101, No. 11534002, No. 11875049, No. U1530402, and No. U1930403), and the National Basic Research Program of China (Grants No. 2016YFA0301201).
\end{acknowledgments}

\bibliographystyle{apsrev}
\bibliography{TSL-ST}

\end{document}